\documentclass[sort&compress]{elsarticle}

\usepackage{color}
\usepackage{amsmath}
\usepackage{amsthm,amscd,amsfonts}
\usepackage{amssymb, upref}
\usepackage{graphicx}
\usepackage[dvips]{epsfig}
\usepackage{epsfig}
\usepackage{caption}

\usepackage{rotating}
\usepackage{amsfonts}

\usepackage{lineno}
\begin{document}
%%%%%%%%%%%%%%%%%%%%%%%%%%%%%%%%%%%%%%%%%%%%%%
\begin{frontmatter}
\title{
Assessing the impact of non-vaccinators: quantifying the average length of infection chains in outbreaks of vaccine-preventable disease
}

\author[st]{S.Towers\corref{cor1}}
\ead{smtowers@asu.edu}
\author[la]{L.J.S. Allen}
\author[fb]{F. Brauer}
\author[st]{B. Espinoza}

\address[st]{Arizona State University, Tempe, AZ, USA}
\address[la]{Texas Tech University, Lubbock, TX, USA}
\address[fb]{University of British Columbia, Vancouver, BC, Canada}

%%%%%%%%%%%%%%%%%%%%%%%%%%%%%%%%%%%%%%%%%%%%%%
%%%%%%%%%%%%%%%%%%%%%%%%%%%%%%%%%%%%%%%%%%%%%%
%%%%%%%%%%%%%%%%%%%%%%%%%%%%%%%%%%%%%%%%%%%%%%

\begin{abstract}
Analytical expressions for the basic reproduction number, ${\cal{R}}_0$, have been obtained in the past for a wide variety of mathematical models for infectious disease spread, along with expressions for the expected final size of an outbreak.  However, what has so far not been studied is the average number of infections that descend down the chains of infection begun by each of the individuals infected in an outbreak (we refer to this quantity as the ``average number of descendant infections'' per infectious individual, or ANDI).  ANDI includes not only the number of people that an individual directly contacts and infects, but also the number of people that those go on to infect, and so on until that particular chain of infection dies out.  Quantification of ANDI has relevance to the vaccination debate, since with ANDI one can calculate the probability that one or more people are hospitalised (or die) from a disease down an average chain of infection descending from an infected un-vaccinated individual.

Here we obtain estimates of ANDI using both deterministic and stochastic modelling formalisms.  With both formalisms we find that even for relatively small community sizes and under most scenarios for ${\cal{R}}_0$ and initial fraction vaccinated, ANDI can be surprisingly large when the effective reproduction number is $>1$, leading to high probabilities of adverse outcomes for one or more people down an average chain of infection in outbreaks of diseases like measles.

\end{abstract}
\end{frontmatter}

%uncomment out for ArXiv
\section{Introduction}

In 1998, Andrew Wakefield published a study that purported a link between
the Mumps/Measles/Rubella (MMR) vaccines and autism~\cite{wakefield1998ileal,wakefield1999mmr}.
The study has been widely debunked~\cite{flaherty2011vaccine,mrozek2010lack,kaye2001mumps,dales2001time}, and has since been retracted by the publisher due to research misconduct and fraud~\cite{fang2012misconduct}.
However, the study has had a significant and continuing
negative impact on attitudes towards vaccination in
developed countries, even though
the American Academy of Pediatrics, the Royal College of Paediatrics and Child Health, the Institute of Medicine, and the World Health Organization have considered the evidence
and endorsed the safety and continuing use of vaccines, including the MMR vaccine~\cite{elliman2002measles}.
Vaccine refusal is rising~\cite{sugerman2010measles,majumder2015substandard}, and delayed or partial vaccination is also becoming an increasing problem~\cite{majumder2015substandard}.

From the perspective of public health policy, the focus tends to be
on the population-level risk of an infectious disease
outbreak due to substandard vaccination coverage,
and the potential number of hospitalisations or deaths that
might result should an outbreak occur~\cite{centers2010national,majumder2015substandard}.

At the individual level however, the anti-vaccine proponents in the vaccination debate 
tend to over-estimate the
risks to the individual should they get vaccinated, while
often under-estimating the risks of
adverse outcomes should they end up catching the disease~\cite{kata2010postmodern,kata2012anti}.
From the individual perspective, 
the argument might indeed be made for some diseases that an
un-vaccinated but healthy
older child or adult might be at somewhat lower risk of adverse outcome should they get the
disease.  However, 
largely overlooked both in the literature and in the vaccination debate is quantification of the impact that an individual's decision not to vaccinate might have on the health outcomes of others in an outbreak situation, particularly for vulnerable individuals in the population like young infants who are too young to be vaccinated; hospitalisations or deaths that, but for that individual's decision to not vaccinate, would potentially not have occurred 
%if it were not for 
downstream in
the infection chain that began with the deliberately un-vaccinated 
individual~\cite{caplan2012free}.

In this analysis we quantify the average number of
infections that are produced down the chain of infection that begins with an individual.
These infections include not only the people that an individual directly infects, but
the people that those go on to infect, and so on until the chain of infection eventually
dies out.
We refer to this quantity as the ``average number of descendant infections'' per infectious
individual, or ANDI.

Quantification of ANDI allows assessment of the risk to others posed by the
infection of an individual.
For example, 
%if the probability that an individual might be hospitalised upon catching a disease is $p$,
if the probability of hospitalisation upon catching a disease is $p$,
the probability that an individual's infection results in the hospitalisation of 
at least one other person
down the chain of infection that began with them
is $1-(1-p)^{\rm ANDI}$.
For large ANDI, this approaches one even for relatively small $p$.
Thus the risk to an individual should they catch the disease might be low, but
the risk may be high that a long chain of infection that began
with that individual resulted in the hospitalisation or death of another.

Here we set up the mathematical and computational formalisms for estimation of
ANDI, using in this introductory work a Susceptible, Infected, Recovered (SIR) model.
We begin by deriving an expression for ANDI from the SIR deterministic model, and compare the results
to those obtained using a stochastic Agent Based Monte Carlo (ABMC) computational approach that keeps
track of who infects whom.
Deterministic modelling methods have the advantage of
computational tractability~\cite{aparicio2007building,rahmandad2008heterogeneity},
especially for large population sizes, but for values of the effective reproduction
number, ${\cal{R}}_{\rm eff}$, close to one, deterministic
models do a poor job of estimation of various quantities associated with an outbreak, including
quantities like
the final size and duration~\cite{greenwood2009stochastic,tritch2018duration}.
%Indeed, as we will show, a key assumption that must be made in the deterministic estimation of ANDI is violated when ${{\cal{R}}_{\rm eff}}$ is close to one and/or for small population sizes. For such cases the ABMC approach is the preferred method for estimation of ANDI.

Using these modelling formalisms, we estimate ANDI for a variety of population sizes, values
of ${\cal{R}}_0$, and pre-outbreak prevalence of vaccinated individuals.

As we will show,
ANDI grows as $\log{N}$, but
ANDI does not
 rise monotonically in ${\cal{R}}_{\rm eff}$.
Rather, relatively small
values of ${\cal{R}}_{\rm eff}$ maximise ANDI,
and even for population sizes representing a small community (for example, $N=10,000$)
ANDI can be surprisingly high, with
each infected individual producing on average one dozen to several dozen descendant infections.
Simulation of a measles-like outbreak in a small community with sub-standard
vaccination prevalence shows that, even in that small population, the probability that
at least one person is hospitalised down an average chain of infection is nearly 100\%.

In the following sections, we describe the deterministic and computational modelling methodology,
followed by a presentation of representative results and discussion.

%%%%%%%%%%%%%%%%%%%%%%%%%%%%%%%%%%%%%%%%%%%%%%%%%%%%%%%%%%%%%%%%%%%%%%%%%%%%%%%%%%
%%%%%%%%%%%%%%%%%%%%%%%%%%%%%%%%%%%%%%%%%%%%%%%%%%%%%%%%%%%%%%%%%%%%%%%%%%%%%%%%%%
%%%%%%%%%%%%%%%%%%%%%%%%%%%%%%%%%%%%%%%%%%%%%%%%%%%%%%%%%%%%%%%%%%%%%%%%%%%%%%%%%%
%%%%%%%%%%%%%%%%%%%%%%%%%%%%%%%%%%%%%%%%%%%%%%%%%%%%%%%%%%%%%%%%%%%%%%%%%%%%%%%%%%
\section{Methods and Materials}

\subsection{The Kermack-McKendrick deterministic SIR epidemic model}
\label{sec:determ}

We begin with a compartmental deterministic model that describes the spread of
a disease in a population that consists of susceptible individuals who
can catch the disease, infectious individuals who may infect them, and recovered and
immune individuals.
The ordinary differential equations describing the dynamics
of the Susceptible, Infected, Recovered (SIR) epidemic model are~\cite{brauer2012mathematical}
\begin{eqnarray} \label{SIR}
S' &=& - \beta\frac{SI}{N} \nonumber \\
I' &=& \beta\frac{SI}{N} - \gamma I\nonumber \\
R' &=& \gamma I.
\label{eqn:sir}
\end{eqnarray}
Here we assume that the dynamics of the outbreak occur on a much
shorter time scale than those of population vital dynamics and the
dynamics of movements in and out of the population,
and thus the population size is constant $N=S+I+R$.

%The basic reproduction number, ${\cal{R}}_0$, is a key quantity that informs outbreak probability of infectious disease;  combined with the population size, levels of pre-outbreak immunity, and the dynamics of the spread of the contagion within a population, it also is related to the final size of an outbreak~\cite{diekmann2000mathematical,allen2008mathematical,brauer2012mathematical}.  In general, for a given disease circulating in a population, the larger the reproduction number, the larger will be the expected final size of the outbreak.

%Analytic expressions for ${\cal{R}}_0$ and the expected outbreak final size have been obtained in the past for an extremely wide variety of dynamical models for the spread of contagious disease (see, for example, the discussion on this topic in References~\cite{diekmann2000mathematical,allen2008mathematical,brauer2012mathematical}).

It is well known 
that the basic reproduction number when the
entire population is na{\"i}ve prior to 
the outbreak is ${\cal{R}}_0=\beta/\gamma$~\cite{brauer2012mathematical}, 
and the effective reproduction number is ${{\cal{R}}_{\rm eff}\!=\!(1\!-\!f^{\rm immune}) {\cal{R}}_0}$, 
where $f^{\rm immune}$
is the fraction immune to the disease before the outbreak.
An outbreak occurs when ${{\cal{R}}_{\rm eff}\!>\!1}$ ~\cite{brauer2012mathematical}.
%and ${S_0\!>\!(1-f^{\rm immune})N}$~\cite{brauer2012mathematical}.
There is also the final size relation when ${\cal{R}}_{\rm eff}\!>\!1$~\cite{miller2012note}
\begin{eqnarray}
%\log\frac{S_0}{S_\infty} = \mathcal{R}_0 \left[1 - \frac{S_\infty}{N}\right].
\log\frac{S_0}{S_\infty} = {{{\cal{R}}_0}\over{N}} \left[(1\!-\!f^{\rm immune})N  - {S_\infty}\right].
\label{eqn:final}
\end{eqnarray}
For a given population size and $f^{\rm immune}$, 
the final size as a fraction of the population, ${(S_0-S_\infty)/N}$, grows
monotonically in ${\cal{R}}_0$. 
\subsubsection{Assessing the average number of descendant infections, ANDI}
\label{sec:assessing}
%\label{sec:limits}

Let ${\rm NDI}(\tau)$ be the number of descendant infections down the infection chain
begun by an individual who was infected at time $\tau$.
Since the total number of new infections occurring after time $\tau$ is $S(\tau) - S_\infty$,
and the fraction of subsequent infections due to a particular individual at time $\tau$ is
on average $1/I(\tau)$\footnote{As an example, when time $\tau=0$, the index case of the outbreak is responsible for 100\% of the subsequent infections.  However, near the peak of an outbreak, when $I(\tau)$ may be, for example, a larger
number like 100, the probability that one of those infectious people is responsible for
any given subsequent infection during the remaining outbreak is 1/100.},
we have
\begin{eqnarray}
{\rm NDI}(\tau) = \frac{S(\tau) - S_\infty}{I(\tau)}.
\label{eqn:M}
\end{eqnarray}
This relation is valid provided $I(\tau) \ge 1$.
%In Methods Section~\ref{sec:limits} we examine the conditions for which this is not the case.
%For $I(\tau)\ge1$, we have ${\rm NDI}(\tau) \le S(\tau) - S_\infty$.

Let $p(\tau)$ be the probability that a susceptible individual is newly infected at time $\tau$.
Since the rate of new infections for
an $SIR$ model is $- S'(\tau)$, and the total number of individuals infected in the outbreak
is $(S_0-S_\infty)$,
we have
\[
p(\tau) = \frac{-S'(\tau)}{S_0-S_\infty}.
\]

The average number of descendant infections caused by all individuals in the outbreak is
the time average of the NDI for infected individuals at time $\tau$, weighted by the probability that the individuals
were newly infected at time $\tau$:
\begin{eqnarray}
{\rm ANDI} &=& \int_0^\infty d\tau \hspace*{1mm} {\rm NDI}(\tau)\; p(\tau) \\
        &=& - \int_0^\infty d\tau \hspace*{1mm} \frac{S'(\tau)}{I(\tau)}
                                              \frac{(S(\tau) - S_\infty)}{(S_0-S_\infty)}
\label{eqn:avg_ANDI}.
\end{eqnarray}

Because the model of Equations~\ref{eqn:sir} is non-linear,
no analytic solution for Equation~\ref{eqn:avg_ANDI} exists. However for
given initial conditions and hypotheses of ${\cal{R}}_0$,
estimates of ANDI can be obtained through numerical integration of the system.

%In the supplementary material, we show that the expected dependence of ANDI on population
%size grows as $\log{N}$, and we also discuss in detail the conditions under which Equation~\ref{eqn:avg_ANDI}
%is expected to be a poor approximation; namely, low ${\cal{R}_0}$ or small population sizes).
%Under these conditions, stochastic modelling methods are preferred for estimation of ANDI.

%%%%%%%%%%%%%%%%%%%%%%%%%%%%%%%%%%%%%%%%%%%%%%
%%%%%%%%%%%%%%%%%%%%%%%%%%%%%%%%%%%%%%%%%%%%%%
\subsubsection{Expected dependence of ANDI on the population size}

Note that Equation~\ref{eqn:avg_ANDI} can be recast as
\begin{eqnarray}
{\rm ANDI} &=& - \int_{S_0}^{S_\infty} dS \hspace*{1mm}
                                              \frac{(S - S_\infty)}{I (S_0-S_\infty)}.
\end{eqnarray}
We note that the integrand has units of one over the population size, $1/N$, and the
measure is proportional to $dN$.  Thus, we expect the integral to grow approximately as $\log{N}$.
This is examined numerically in Results Section~\ref{sec:logN}.

%%%%%%%%%%%%%%%%%%%%%%%%%%%%%%%%%%%%%%%%%%%%%%
%%%%%%%%%%%%%%%%%%%%%%%%%%%%%%%%%%%%%%%%%%%%%%
\subsection{Limitations of the deterministic approach}
\label{sec:limits}

In the derivations presented in Methods Section~\ref{sec:determ},
note that Equation~\ref{eqn:M} only holds
when $I(\tau)\!\ge\!1$.
The possibility that $I(\tau)\!<\!1$
is problematic if the number of susceptible individuals still left to infect, $x\!=\!S(\tau)\!-\!S_\infty$,
is also greater
than one (particularly if it is much greater than one);
in this case, effects due to population stochasticity will be non-negligible in the
overall outbreak dynamics, and the deterministic approximation will likely be poor.
The deterministic approximation will also be poor for small population sizes, because
population stochasticity will again be non-negligible in that case.

To determine the conditions for which $x$ is large when ${I\!<\!1}$, we begin by solving for 
the phase curve, $I(S)$, obtained from the solution of the Equations~\ref{eqn:sir}.
Equations~\ref{eqn:sir} can be recast
as
\begin{eqnarray}
dI/dS & = & {{\beta S I/N - \gamma I}\over{-\beta S I/N}} \nonumber \\
      & = & {{{1}\over{{\cal{R}}_0}}{{N}\over{S}}-1}.
\label{eqn:dIdS}
\end{eqnarray}
Integrating Equation~\ref{eqn:dIdS} with respect to $S$ yields
\begin{eqnarray}
I = {{N}\over{{\cal{R}}_0}} \log{S} - S + c.
\label{eqn:I}
\end{eqnarray}
For an outbreak with initial conditions $I\!=\!I_0$ and ${S\!=\!S_0\!=\!(1\!-\!f^{\rm immune})N\!-\!I_0}$,
we see that
Equation~\ref{eqn:I} is satisfied when
\begin{eqnarray}
c & = & S_0 + I_0 - {{N}\over{{\cal{R}}_0}} \log{S_0} \nonumber \\
  & = & (1\!-\!f^{\rm immune})N + {{N}\over{{\cal{R}}_0}} \log{{{1}\over{S_0}}}.
\end{eqnarray}
Substituting this into Equation~\ref{eqn:I} yields the phase relation
\begin{eqnarray}
  I(S) &=& \frac{N}{{\cal{R}}_0} \log{\left( \frac{S}{S_0} \right)} + (1\!-\!f^{\rm immune})N-S.
\label{eqn:iphase}
\end{eqnarray}
%If $x$ is the number of individuals left to be infected at some point in the outbreak (out
%of the total infected in the entire outbreak), then
Now
when the number still left to infect is $x$, we have
${S\!=\!S_\infty+x}$.  We thus have
\begin{eqnarray}
  I &=& \frac{N}{{\cal{R}}_0} \log{\left( \frac{S_\infty+x}{S_0} \right)} + (1\!-\!f^{\rm immune})N-S_\infty-x.
\label{eqn:iphaseb}
\end{eqnarray}
Note that we can recast this as
\begin{eqnarray}
I & = & \frac{N}{{\cal{R}}_0} \log{\left( \frac{S_\infty}{S_0} \right)}
       + \left[(1\!-\!f^{\rm immune})N-S_\infty\right]\nonumber \\
   & &   + \frac{N}{{\cal{R}}_0} \log{\left( 1+\frac{x}{S_\infty} \right)}
      -x.
\label{eqn:iphasec}
\end{eqnarray}
We recognise the first two terms as the two sides of the final size relation in Equation~\ref{eqn:final},
multiplied by $N/{\cal{R}}_0$.  The first two terms thus cancel, yielding
\begin{eqnarray}
I & = &
      \frac{N}{{\cal{R}}_0} \log{\left( 1+\frac{x}{S_\infty} \right)}
      -x.
\label{eqn:iphased}
\end{eqnarray}
When $I=1$, we obtain
\begin{eqnarray}
      \frac{N}{{\cal{R}}_0} \log{\left( 1+\frac{x}{S_\infty} \right)}
      -x
& = & 1.
\label{eqn:iphasee}
\end{eqnarray}
Now, for $u\ge0$, $u\ge\log(1+u)$\footnote{Proof: Consider the function $g(u) = \log(1+u)-u$.
Then $g(u)=0$ and $g^\prime(u)=1/(1+u)-1\le0$ for $u\ge0$.  Thus the maximum of $g(u)$ for $u\ge0$ is
$g(0)=0$.}, thus from Equation~\ref{eqn:iphasee} we obtain
\begin{eqnarray}
      \frac{N}{{\cal{R}}_0} \frac{x}{S_\infty}
      -x
&\ge& 1.
\label{eqn:iphasef}
\end{eqnarray}
Solving for $x$ thus yields
\begin{eqnarray}
x \ge {{{\cal{R}}_0 S_\infty/N}\over
     {1-{\cal{R}}_0 S_\infty/N}}.
\label{eqn:xlim}
\end{eqnarray}
As ${{\cal{R}}_{\rm eff}\!\rightarrow\!1}$, 
%the outbreak final size
%shrinks to zero (i.e. ${S_\infty\!\rightarrow\!S_0}$), and
we see from Equation~\ref{eqn:final} that 
${S_\infty{{\cal{R}}_0}/N\!\rightarrow\!1}$.
%Because $S_0/S_\infty>1$ when ${\cal{R}}_{\rm eff}>1$, we see from Equation~\ref{eqn:final}
%that ${{\cal{R}}_0} S_\infty/N<{{\cal{R}}_{\rm eff}}$.
%Recall that $S_\infty/N\!<\!1/{\cal{R}}_0$, thus the RHS of the expression is greater than zero,and also implies that 
%$x\!\ge\!{{\cal{R}}_0}/(1-{\cal{R}}_0)$.
%We see that as ${\cal{R}}_0\!\rightarrow\!1$,
Thus, as ${\cal{R}}_{\rm eff}\!\rightarrow\!1$,
we see from Equation~\ref{eqn:xlim} that by the time $I(\tau)$ falls to one the
number still left to infect in the outbreak, ${x\!=\!S(\tau)-S}$, can be large,
% (violating the assumption of Equation~\ref{eqn:M}), 
and thus the deterministic formalism in such cases will yield a poor description of the
evolution of the SIR system.

%%%%%%%%%%%%%%%%%%%%%%%%%%%%%%%%%%%%%%%%%%%%%%%%%%%%%%%%%%%%%%%%%%%%%%%%%%%%%%%%%%
%%%%%%%%%%%%%%%%%%%%%%%%%%%%%%%%%%%%%%%%%%%%%%%%%%%%%%%%%%%%%%%%%%%%%%%%%%%%%%%%%%
%%%%%%%%%%%%%%%%%%%%%%%%%%%%%%%%%%%%%%%%%%%%%%%%%%%%%%%%%%%%%%%%%%%%%%%%%%%%%%%%%%
%%%%%%%%%%%%%%%%%%%%%%%%%%%%%%%%%%%%%%%%%%%%%%%%%%%%%%%%%%%%%%%%%%%%%%%%%%%%%%%%%%
\subsection{Stochastic Modelling Methods}
\label{sec:stoch}

To estimate not only the expected value of ANDI, but also the range in that
quantity due to population stochasticity, we used an 
ABMC model, since it is
only this stochastic modelling formalism that allows us to keep track of who infects whom at the individual level.

ABMC simulations of epidemics
involve setting up probabilistic and/or heuristic ``rules'' 
for the interactions of individuals in the population, and the sojourn times spent
%in the disease states~\cite{stroud2007spatial,helbing2012agent}.
in the disease states~\cite{helbing2012agent}.
In the ABMC simulation for the dynamics of an SIR model with homogeneous mixing,
%we assumed that an individual was equally likely at a particular time step to contact
%any other individual in a population.
at each time step of length $\Delta \tau$ (in units of $1/\gamma$),
we determined if an infectious individual recovered, with average probability
$p^{\rm recover} = (1-e^{-\Delta \tau})$.  If a sampled uniformly distributed random
number was less than $p^{\rm recover}$, the individual was moved to the recovered class.

Additionally, at each time step we calculated the expected number of contacts that
each susceptible person made with infectious people, $\beta I \Delta\tau/N$, and sampled a
Poisson distributed number with this mean. To emulate homogeneous mixing, we assumed
that there was no preferential mixing of the population, and each individual was equally
likely to contact any other individual in the population during a particular time step.
If the sampled number of contacts with infectious
people was greater than zero, the susceptible
individual was moved to the infectious class.
An individual was randomly sampled from the list of infectious individuals as the
parent of this new infection.
% (again, because of the homogeneous mixing ansatz, it
%did not matter which infectious individual was selected as the parent).

For each infected individual we kept track of the time at which they were infected,
who infected them, and also who they subsequently infected.
From this information, we calculated the number of infected descendants
of each individual, and the average and range of ANDI for all individuals infected in the outbreak.
%For each outbreak simulation, the average number of descendant infections were determined, as well as the 95\% confidence upper limit on the number of desendant infections.

Due to population stochasticity,
not all outbreaks progress~\cite{allen2010introduction,allen2013relations,house2013how,black2015computation}; 
indeed, the probability that no individuals are infected
beyond the first initial infection is ${1/(1+{\cal{R}}_{\rm eff})}$~\cite{whittle1955outcome,bailey1975mathematical}.
There has also been a distinction made between ``minor'' outbreaks where just a few
individuals are infected before the outbreak sputters out, and ``major'' outbreaks where
the final size is much closer to the deterministic prediction~\cite{whittle1955outcome}.  
Here we take a conservative
approach, and estimate ANDI from both minor and major outbreaks in the ABMC simulations;
this will underestimate ANDI compared to estimation using only major outbreaks, and thus in essence
represents the ``best case scenario'' assessment of risk to the population posed by un-vaccinated
individuals.

To cross-check the fidelity of the ABMC simulations, we also employed a
continuous-time Markov chain (CTMC) SIR model~\cite{allen2010introduction}, 
and compared the final size distributions
given by the two simulation methods to ensure that they were consistent
(note that the CTMC model cannot track who infects whom).

For even moderate population sizes and even modestly large ${\cal{R}}_{\rm eff}$, the
ABMC simulations were computationally intensive, and required the use of high performance
computing resources to achieve a large number of stochastic realisations of the system.
The simulations in this analysis were performed
using the high performance computing platforms at Arizona State University
and Texas Tech University.
The authors have produced a library for the
\textbf{\textsf{R}} statistical programming language that contains functions that
perform both the deterministic and stochastic modelling methods applied in this analysis.
The \textbf{\textsf{R}} {\tt{ANDI}} library can be downloaded from the
GitHub repository {\tt https://github.com/smtowers/ANDI}.
To do this at the \textbf{\textsf{R}} command line, type
\begin{verbatim}
  require("devtools")
  install_github("smtowers/ANDI")
  require("ANDI")
\end{verbatim}
An example \textbf{\textsf{R}} script showing the use of the methods can be found at
{\tt https://github.com/smtowers/ANDI/example.R}.

%%%%%%%%%%%%%%%%%%%%%%%%%%%%%%%%%%%%%%%%%%%%%%%%%%%%%%%%%%%%%%%%%%%%%%%%%%%%%%%%%%
%%%%%%%%%%%%%%%%%%%%%%%%%%%%%%%%%%%%%%%%%%%%%%%%%%%%%%%%%%%%%%%%%%%%%%%%%%%%%%%%%%
%%%%%%%%%%%%%%%%%%%%%%%%%%%%%%%%%%%%%%%%%%%%%%%%%%%%%%%%%%%%%%%%%%%%%%%%%%%%%%%%%%
%%%%%%%%%%%%%%%%%%%%%%%%%%%%%%%%%%%%%%%%%%%%%%%%%%%%%%%%%%%%%%%%%%%%%%%%%%%%%%%%%%
\subsection{Modelling scenarios examined}

We used both the deterministic and stochastic formalisms to obtain estimates of ANDI and the outbreak
final size for values of ${\cal{R}}_0$ between $1.1$ to $4$ in steps of $0.1$, under the assumption that the
initial population was entirely susceptible except for one infectious individual.
We did this for population sizes $N=1,000$, $N=2,500$, $N=5,000$, $N=7,500$ and $N=10,000$, and
also for $N=100,000$ with a smaller selection of 
representative values of ${\cal{R}}_0$.
For each value of ${\cal{R}}_0$ and $N$, we performed at least $5,000$ stochastic realisations.
The stochastic modelling formalism was
computationally intensive and the complexity
grows non-linearly in $N$;
thus, due to the wide variety of reproduction number values we examined, it was
computationally infeasible
to examine larger population sizes in this analysis.
However, with the population sizes we did examine, we
were able to estimate the apparent dependence of ANDI on $N$, ${\cal{R}}_0$, and
the final size.

The one group SIR model we employed in this analysis
did not explicitly include a vaccinated class, but we
were nevertheless able to examine scenarios that involved a fraction of the population that was effectively
vaccinated and fully
immune to the disease prior to the outbreak by including those individuals in the initial conditions for the
``recovered'' class.
An example of a disease for which this is a good approximation is measles,
for which the vaccine is known to have an efficacy in excess of 95\%~\cite{majumder2015substandard,janaszek2003measles,mossong2000estimation,van2010estimation}, usually
conferring life-long humoral immunity~\cite{amanna2007duration}.
Measles has a high basic reproduction number
in the absence of vaccination, between $11$ to $18$~\cite{plans2012evaluation,majumder2015substandard}. Thus, even though vaccination rates usually exceed 90\% in most
areas of the US, the effective reproduction number is still
high enough that outbreaks are possible~\cite{majumder2015substandard}.
%We thus estimated the values of ANDI for a hypothetical measles outbreak with a basic reproduction
%number of ${\cal{R}}_0=15$ in a 90\% vaccinated population under the assumption of 100\% vaccine
%efficacy.
%The un-vaccinated individuals in the population consist of deliberately un-vaccinated
%individuals,
%plus the approximately $1.2\%$ of the overall population that consists of babies under the
%age of one year old who are too young to be
%%vaccinated\footnote{Census data on the U.S. population
%%in one year age groups can be downloaded from the National Cancer Institute
%%{\tt https://seer.cancer.gov/popdata/download.html}, accessed December, 2018.}.
%vaccinated\footnote{The 2017 census
%data on the U.S. population in one year age group can be downloaded from
%%{\tt https://www.census.gov/data/tables/2017/demo/popest/nation-detail.html},
%{\tt https://bit.ly/2CEtI8W},
%accessed December, 2018.}.

Using the deterministic and stochastic modelling formalisms,
we thus simulated a measles-like
outbreak with basic reproduction number ${\cal{R}}_0=15$ in a population size of $N=10,000$,
examining various scenarios where the pre-immune fraction, $f^{\rm immune}$, was sampled
from $0$ to $1$ in steps of $0.1$.

%%%%%%%%%%%%%%%%%%%%%%%%%%%%%%%%%%%%%%%%%%%%%%%%%%%%%%%%%%%%%%%%%%%%%%%%%%%%%%%%%%
%%%%%%%%%%%%%%%%%%%%%%%%%%%%%%%%%%%%%%%%%%%%%%%%%%%%%%%%%%%%%%%%%%%%%%%%%%%%%%%%%%
%%%%%%%%%%%%%%%%%%%%%%%%%%%%%%%%%%%%%%%%%%%%%%%%%%%%%%%%%%%%%%%%%%%%%%%%%%%%%%%%%%
\section{Results}

%%%%%%%%%%%%%%%%%%%%%%%%%%%%%%%%%%%%%%%%%%%%%%
%%%%%%%%%%%%%%%%%%%%%%%%%%%%%%%%%%%%%%%%%%%%%%
\subsection{Relationship between ANDI and ${\cal{R}}_0$ and the final size}

\begin{figure}[h]
   \begin{center}
      %\includegraphics{file=sir_agent_determ_fig_N1000.eps, width=12cm}
      %\includegraphics[width=0.90\linewidth]{sir_agent_determ_fig_N1000}
      %\epsfig{file=sir_agent_determ_fig_N1000.eps, width=0.9\linewidth}
      %\vspace*{-3.5cm}
      \epsfig{file=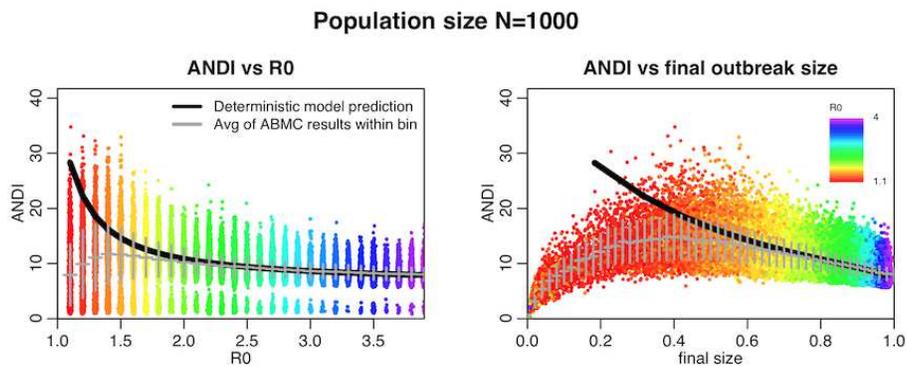, width=1.0\linewidth}
      \vspace*{-0.5cm}
  \caption{
      \label{fig:agent1000}
   The left hand plot 
   shows the average number of descendant infections,
   ANDI, versus 
   the basic reproduction number, ${\cal{R}}_0$,
   for $N=1,000$ and various hypotheses of ${\cal{R}}_0$ between $1.1$ to $4$,
   as estimated by the ABMC stochastic SIR modelling formalism described in
   Methods Section~\ref{sec:stoch}.
   The right hand plot 
   shows ANDI versus 
   %${\cal{R}}_0$.  
   the final size of the outbreak.
   %Only outbreaks which progressed and did not die out early
   %are shown in this plot; that is to say outbreaks with final size nominally greater than 10\%.
   Overlaid are the predictions derived from
   the deterministic SIR model, described in Methods Section~\ref{sec:determ}.
   The height of the grey bars represents the one standard
   deviation variation of the ABMC results within 
   bins of ${\cal{R}}_0$,
   or 
   bins of final size. 
   Note that the slight side-to-side scatter in the left hand plot is for
   clarity of display purposes only; each point corresponds to a stochastic realisation of the 
   ABMC simulation at an exact value of ${\cal{R}}_0$.
   %As described in the text, it is only for larger values of ${\cal{R}}_0$ that the deterministic estimates are good approximations.
   }
   \end{center}
 \end{figure}

\begin{figure}[h]
   \begin{center}
      %\includegraphics{file=sir_agent_determ_fig_N10000.eps, width=12cm}
      %\includegraphics[width=1.00\linewidth]{sir_agent_determ_fig_N10000}
      %\epsfig{file=sir_agent_determ_fig_N10000.eps, width=0.9\linewidth}
      %\vspace*{-3.5cm}
      \epsfig{file=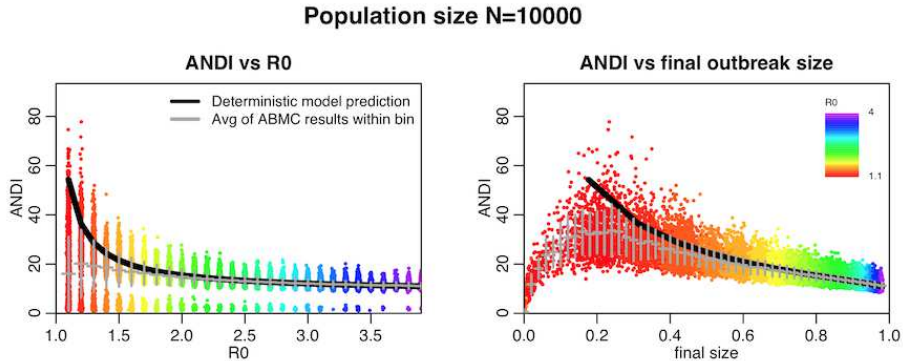, width=1.0\linewidth}
      \vspace*{-0.5cm}
  \caption{
      \label{fig:agent10000}
   As Figure~\ref{fig:agent1000}, but with population size $N=10,000$.
   }
   \end{center}
 \end{figure}

In Figure~\ref{fig:agent1000} we show the results of the deterministic
and ABMC simulations, showing
the predicted ANDI versus ${\cal{R}}_0$, and versus the final size of the outbreak,
when the population size is $N=1,000$ and for the various hypotheses of
${\cal{R}}_0$ for outbreaks occurring in entirely na{\"i}ve populations.
In Figure~\ref{fig:agent10000} we show the results
when the population size is $N=10,000$ and for the various hypotheses of
${\cal{R}}_0$.
For values of ${\cal{R}}_0\gtrapprox1.5$ the deterministic and 
ABMC modelling
results are in agreement.  For small values of ${\cal{R}}_0$ the
predictions diverge, 
%due to the reasons discussed in Methods Section~\ref{sec:limits}, 
and the ABMC model is preferred.

For an outbreak in a small community with population $N=10,000$,
the ABMC results indicate that ANDI is maximised 
for middling values of ${\cal{R}}_0$ around $1.2$ to $1.5$,
and for outbreak final sizes around 20\%.

%We note that for larger ${\cal{R}}$ (${\cal{R}}_0\gtrapprox1.5$) and a given population size, ANDI has an approximately negative linear relationship with the final size of the outbreak.

%%%%%%%%%%%%%%%%%%%%%%%%%%%%%%%%%%%%%%%%%%%%%%
%%%%%%%%%%%%%%%%%%%%%%%%%%%%%%%%%%%%%%%%%%%%%%
\subsection{Relationship between ANDI and the population size}
\label{sec:logN}

\begin{figure}[h]
   \begin{center}
      \epsfig{file=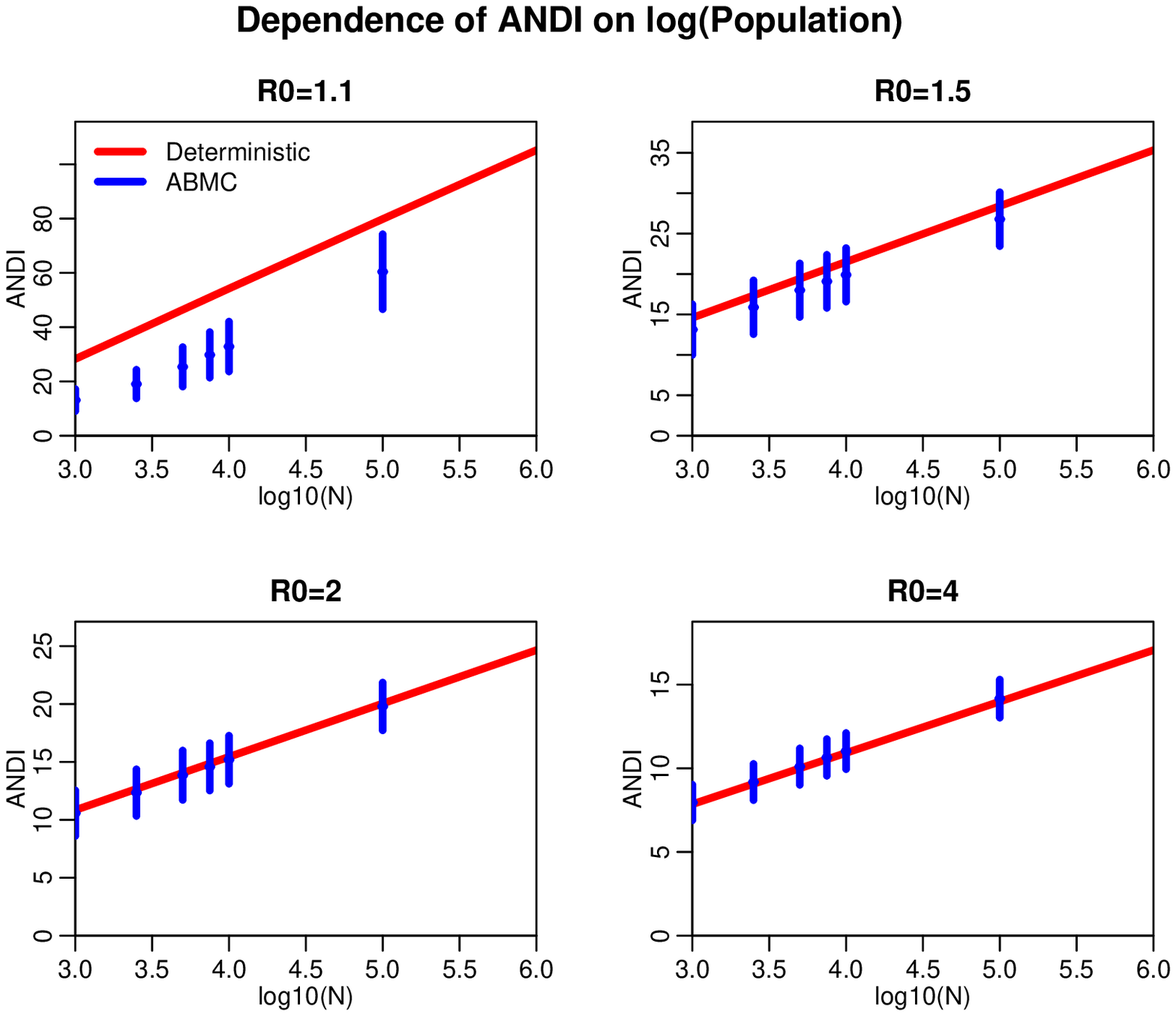, width=0.9\linewidth}
  \vspace*{-0.5cm}
  \caption{
      \label{fig:logn}
   Variation of ANDI versus the logarithm of the population size, as predicted
   by the deterministic and ABMC models for various values of the reproduction
   number.  The disagreement in the values for small ${\cal{R}}_0$ are for the
   reasons discussed in Methods Section~\ref{sec:limits}.
    The vertical bars on the ABMC model results indicate the one standard
    deviation variation in the simulation results.
   }
   \end{center}
 \end{figure}

In Figure~\ref{fig:logn} we show the variation of ANDI versus the
logarithm of the population size for the deterministic and 
ABMC simulations
under a range of values of the reproduction number.
%We note that the relationship between the quantities is, to a good approximation, linear.
%We note that the relationship between ANDI and $\log{N}$ appears to be linear.
Both the deterministic and ABMC modelling results
for ANDI show a linear
dependence on $\log{N}$, but for small values of ${\cal{R}}_0$ the disagreement
between the ABMC and deterministic intercept terms is due to the poor approximation of the
deterministic model under those conditions, for the reasons discussed in Methods Section~\ref{sec:limits}.

%%%%%%%%%%%%%%%%%%%%%%%%%%%%%%%%%%%%%%%%%%%%%%%%%%%%%%%%%%%%%%%%%%%%%%%%
%%%%%%%%%%%%%%%%%%%%%%%%%%%%%%%%%%%%%%%%%%%%%%%%%%%%%%%%%%%%%%%%%%%%%%%%
%%%%%%%%%%%%%%%%%%%%%%%%%%%%%%%%%%%%%%%%%%%%%%%%%%%%%%%%%%%%%%%%%%%%%%%%
% made with sir_agent_determ_fig.R
%%%%%%%%%%%%%%%%%%%%%%%%%%%%%%%%%%%%%%%%%%%%%%%%%%%%%%%%%%%%%%%%%%%%%%%%
%\begin{figure}%[tbhp]
%\centering
%%\includegraphics[width=.8\linewidth]{frog}
%\caption{Placeholder image of a frog with a long example caption to show justification setting.}
%\label{fig:frog}
%\end{figure}

\vspace*{-0.0cm}

%%%%%%%%%%%%%%%%%%%%%%%%%%%%%%%%%%%%%%%%%%%%%%%%%%%%%%%%%%%%%%%%%%%%%%%%
%%%%%%%%%%%%%%%%%%%%%%%%%%%%%%%%%%%%%%%%%%%%%%%%%%%%%%%%%%%%%%%%%%%%%%%%
%%%%%%%%%%%%%%%%%%%%%%%%%%%%%%%%%%%%%%%%%%%%%%%%%%%%%%%%%%%%%%%%%%%%%%%%
% made with sir_agent_determ_fig.R
%%%%%%%%%%%%%%%%%%%%%%%%%%%%%%%%%%%%%%%%%%%%%%%%%%%%%%%%%%%%%%%%%%%%%%%%
\vspace*{-0.5cm}
%%%%%%%%%%%%%%%%%%%%%%%%%%%%%%%%%%%%%%%%%%%%%%%%%%%%%%%%%%%%%%%%%%%%%%%%
%%%%%%%%%%%%%%%%%%%%%%%%%%%%%%%%%%%%%%%%%%%%%%%%%%%%%%%%%%%%%%%%%%%%%%%%
%%%%%%%%%%%%%%%%%%%%%%%%%%%%%%%%%%%%%%%%%%%%%%%%%%%%%%%%%%%%%%%%%%%%%%%%

%%%%%%%%%%%%%%%%%%%%%%%%%%%%%%%%%%%%%%%%%%%%%%%%%%%%%%%%%%%%%%%%%%%%%%%%
%%%%%%%%%%%%%%%%%%%%%%%%%%%%%%%%%%%%%%%%%%%%%%%%%%%%%%%%%%%%%%%%%%%%%%%%
%%%%%%%%%%%%%%%%%%%%%%%%%%%%%%%%%%%%%%%%%%%%%%%%%%%%%%%%%%%%%%%%%%%%%%%%
% made with andi_vs_population.R
%%%%%%%%%%%%%%%%%%%%%%%%%%%%%%%%%%%%%%%%%%%%%%%%%%%%%%%%%%%%%%%%%%%%%%%%
\vspace*{-0.5cm}
%%%%%%%%%%%%%%%%%%%%%%%%%%%%%%%%%%%%%%%%%%%%%%%%%%%%%%%%%%%%%%%%%%%%%%%%
%%%%%%%%%%%%%%%%%%%%%%%%%%%%%%%%%%%%%%%%%%%%%%%%%%%%%%%%%%%%%%%%%%%%%%%%
%%%%%%%%%%%%%%%%%%%%%%%%%%%%%%%%%%%%%%%%%%%%%%%%%%%%%%%%%%%%%%%%%%%%%%%%

%%%%%%%%%%%%%%%%%%%%%%%%%%%%%%%%%%%%%%%%%%%%%%
%%%%%%%%%%%%%%%%%%%%%%%%%%%%%%%%%%%%%%%%%%%%%%
\subsection{Relationship between ANDI and pre-immunity of the population}

\begin{figure}[h]
   \begin{center}
      \epsfig{file=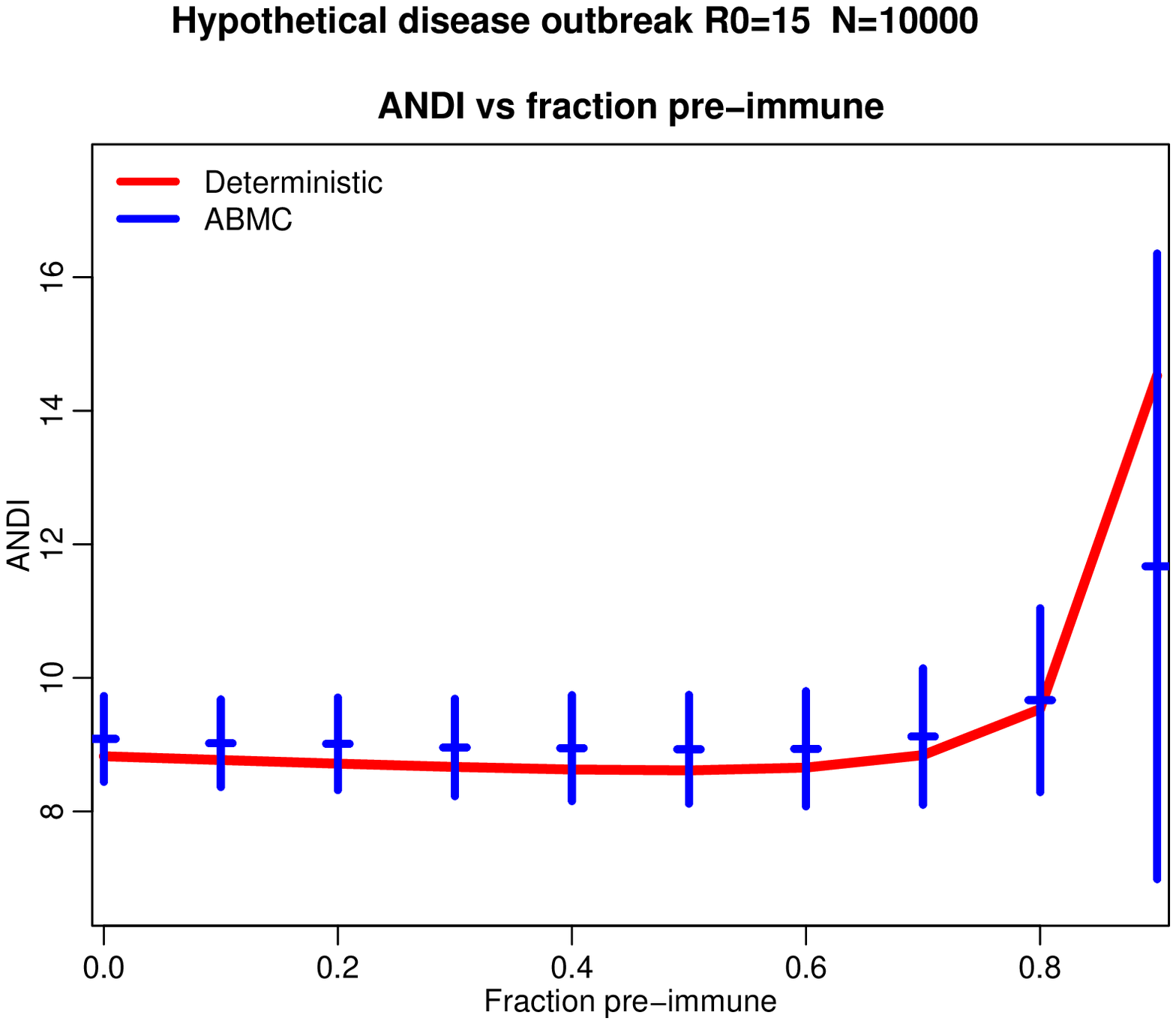, width=0.7\linewidth}
  \vspace*{-0.5cm}
  \caption{
      \label{fig:measles}
   The average number of descendant infections, ANDI,
   versus the fraction of the population with pre-immunity to the disease
   for a hypothetical measles-like
   outbreak with ${\cal{R}}_0=15$ in a population of $N=10,000$
   individuals.
    The vertical bars on the ABMC model results indicate the one standard
    deviation variation in the simulation results.
   %The left hand plot shows the final size of the outbreak versus the pre-immune fraction.
   }
   \end{center}
 \end{figure}

Using the deterministic and stochastic modelling formalisms,
we simulated an SIR
outbreak with basic reproduction number ${\cal{R}}_0=15$ in a population size of $N=10,000$,
and with hypotheses of the pre-immune fraction, $f^{\rm immune}$, ranging from $0$ to $1$ in steps of $0.1$.
The results showing ANDI versus the pre-immune fraction are shown in
Figure~\ref{fig:measles}.
%Approximately $1.2\%$ of the population consists of infants too young to be vaccinated for measles.  When 10\% of the population is un-vaccinated (i.e. ${f^{\rm immune}\!=\!0.9}$), $12\%$ of that un-vaccinated population thus consists of infants.

From Figure~\ref{fig:measles}, we note that
ANDI is almost constant for low to middling values of $f^{\rm immune}$, but rises as $f^{\rm immune}$ approaches the
point where ${\cal{R}}_{\rm eff}$ becomes close to one.  This emulates the situation in
some developed countries where, in some locales, vaccination coverage has dropped just below
the level where measles outbreaks are now possible.

From Figure~\ref{fig:measles}, we note that
in small community of population size $N=10,000$, the ABMC simulations
indicate that ANDI is approximately 12 when ${f^{\rm immune}\!=\!0.9}$ 
for our hypothetical measles outbreak (95\% CI $[1.50,19.8]$).
For larger population sizes, this will grow as $\log{N}$.
%This means that on average ${12\ast0.12=1.35}$ of the individuals down the chain of infection from an un-vaccinated infectious individual were infants.
The case hospitalisation rate of measles is high, with post-outbreak analyses
estimating it to be between $16-43\%$~\cite{van2002measles,dominguez2008large,zipprich2015measles,grammens2017ongoing}.
The U.S. Centers for Disease Control (CDC) estimates on average 25\% of cases require
hospitalisation, and approximately $0.2\%$ result in death (see {\tt https://www.cdc.gov/measles/about/complications.html}, accessed December, 2018).
From the ABMC simulation, the estimated
probability that at least one person in an average descendant chain of infection
in a population size of $N=10,000$ is hospitalised in this hypothetical measles outbreak with ${f^{\rm immune}\!=\!0.9}$
is thus ${1-(1-0.25)^{12}=0.97}$ (95\% CI $[0.35,1.00]$),
and the estimated probability that at least one person in an average descendant chain of infection dies is ${1-(1-0.002)^{12}=0.024}$ (95\% CI $[0.003,0.039]$).
%, and the probability that at least one infant in the average descendant chain of infection is hospitalised is ${1-(1-0.25)^{1.6}=0.37}$.
%probability that at least one infant in the average descendant chain of infection
%dies is ${1-(1-0.002)^{1.6}=0.003}$.

%%%%%%%%%%%%%%%%%%%%%%%%%%%%%%%%%%%%%%%%%%%%%%%%%%%%%%%%%%%%%%%%%%%%%%%%
%%%%%%%%%%%%%%%%%%%%%%%%%%%%%%%%%%%%%%%%%%%%%%%%%%%%%%%%%%%%%%%%%%%%%%%%
%%%%%%%%%%%%%%%%%%%%%%%%%%%%%%%%%%%%%%%%%%%%%%%%%%%%%%%%%%%%%%%%%%%%%%%%
% made with measles.R
%%%%%%%%%%%%%%%%%%%%%%%%%%%%%%%%%%%%%%%%%%%%%%%%%%%%%%%%%%%%%%%%%%%%%%%%
%\vspace*{-0.5cm}
%%%%%%%%%%%%%%%%%%%%%%%%%%%%%%%%%%%%%%%%%%%%%%%%%%%%%%%%%%%%%%%%%%%%%%%%
%%%%%%%%%%%%%%%%%%%%%%%%%%%%%%%%%%%%%%%%%%%%%%%%%%%%%%%%%%%%%%%%%%%%%%%%

%%%%%%%%%%%%%%%%%%%%%%%%%%%%%%%%%%%%%%%%%%%%%%%%%%%%%%%%%%%%%%%%%%%%%%%%
%%%%%%%%%%%%%%%%%%%%%%%%%%%%%%%%%%%%%%%%%%%%%%%%%%%%%%%%%%%%%%%%%%%%%%%%
%%%%%%%%%%%%%%%%%%%%%%%%%%%%%%%%%%%%%%%%%%%%%%%%%%%%%%%%%%%%%%%%%%%%%%%%
\section{Discussion and Summary}

In this introductory work, we for the first time have quantified the average number
of descendant infections, ANDI, that spread down the chains of infection that
begin with individuals infected in an outbreak.  Quantification of ANDI
is necessary to quantify the risk to others posed by the infection of un-vaccinated
individuals in an outbreak of a vaccine-preventable disease; the higher the ANDI,
the more likely at least one person down an average chain of infection is hospitalised or dies.

%We found that ANDI can be remarkably
%high for even modestly sized communities, where average infection chains can consist of many
%people for some values of ${\cal{R}}_{\rm eff}$.

Unlike outbreak final size, which grows linearly in population size, $N$, and grows 
monotonically in ${\cal{R}}_{\rm eff}$,
we found that ANDI grows as $\log{N}$, and is
larger for relatively small values of ${\cal{R}}_{\rm eff}$ rather than for large values,
and in such cases ANDI can be remarkably high, with average infection chains consisting of
many people even in modestly sized communities.
The fact that ANDI does not grow in ${\cal{R}}_{\rm eff}$ can be conceptualised
as follows;
the final size grows monotonically in ${\cal{R}}_{\rm eff}$, but because of the
rapid proliferation of the number infected early in a high ${\cal{R}}_{\rm eff}$ 
outbreak, there are many more infected
individuals subsequently ``sharing the pie''
of those left to infect in outbreak.
%Thus, while at any one time point the remaining
%``pie'' of susceptibles left to infect might be somewhat bigger as ${\cal{R}}_{\rm eff}$ grows,
%there are more individuals sharing it, and each thus gets a smaller slice.
%Another way to conceptualise the effect is to note that high ${\cal{R}}_{\rm eff}$ outbreaks quickly burn through a population and the effective reproduction number quickly degrades, and thus the chains of infection in such outbreaks, while perhaps containing on average more individuals in the first generation of infections in the chain (particularly for those infected early on in the outbreak), will die out quicker than chains of infection in lower ${\cal{R}}_{\rm eff}$ outbreaks.
However, 
%ANDI is not maximised as ${{\cal{R}}_{\rm eff}\!\rightarrow\!1}$, 
our stochastic simulations showed that as ${{\cal{R}}_{\rm eff}\!\rightarrow\!1}$,
effects due to population stochasticity begin to dominate, and the chains of infection
tend to die out quicker, leading to a reduction
in ANDI.
The apparent dependence of ANDI on $\log{N}$ is similar to the $\log{N}$ dependence of the average path length in
random networks~\cite{fronczak2004average}.  Further study is needed to elucidate the
potential reasons for this.

While the model examined in this introductory work was a 
Susceptible, Infected, Recovered model without a specific ``vaccinated''
compartment, it can nevertheless be used
to simulate outbreaks in populations where a fraction of the population has immunity
to the disease prior to the outbreak
(for example, because they had been vaccinated with a highly effective vaccine).
In our analysis
we simulated a hypothetical measles-type outbreak in a relatively small community
of $N=10,000$ people, assuming some fraction
of the population had prior immunity due to completely effective vaccination.  
When the fraction of pre-immune individuals in the population
is just below the limit where an outbreak can occur, with
effective reproduction number ${\cal{R}}_{\rm eff}\sim 1.5$ (similar to the estimated
${\cal{R}}_{\rm eff}$ in observed measles outbreaks in populations with just-substandard vaccine
coverage~\cite{mossong2000estimation}),
we found that ANDI is approximately a dozen
individuals.
%Recall that if the probability that an individual with the disease is hospitalised is $p$, then the probability that at least one individual down a chain of descendant infections is hospitalised is $1-(1-p)^{\rm ANDI}$.
%Because measles has a relatively high case hospitalisation rate,
% of, on average, $p=25\%$, 
Because ANDI is large, 
we found that the probability that at least one individual
 gets hospitalised down an average descendant infection
is close to one in this hypothetical outbreak, even for this modest population size.
The probability will grow even closer to one
for larger population sizes, due to the dependence of ANDI on $\log{N}$.
%Beyond the human cost, the economic cost of these hospitalisations in the descendant infection chains is very high; the CDC estimates that each measles hospitalisation costs between \$4,000 to \$46,000~\cite{whitney2014benefits}.
Because the average descendant chain of infection can be remarkably long, the un-vaccinated
individual may likely not even directly know the person(s) hospitalised down their
infection chain, and it is this distance of association that can lead to under-estimation
of the impact of non-vaccination on others.

It is important to note that infants under the age of one year are too young to be vaccinated
for most infectious diseases, and also tend to be at highest risk of hospitalisation upon
catching such diseases.  
%The hospitalised individuals down a chain of infection in
%this hypothetical outbreak thus include
%not only purposely un-vaccinated individuals, but also infants too young to be vaccinated.
Approximately $1.2\%$ of the population consists of infants too young to be vaccinated for
measles\footnote{The 2017 census
data on the U.S. population in one year age group can be downloaded from
%%{\tt https://www.census.gov/data/tables/2017/demo/popest/nation-detail.html},
{\tt https://bit.ly/2CEtI8W},
accessed December, 2018.};
thus,
when 5\% to 10\% of the population is un-vaccinated (for example),
a relatively large fraction of that un-vaccinated population consists of infants.
Indeed, in a recent outbreak of measles in California,
%(a state where the MMR vaccine coverage for children over 19 months is over 92\%),
it was observed that almost 25\%
of the cases in un-vaccinated individuals occurred in infants too young to be vaccinated~\cite{zipprich2015measles}.
Thus, while individuals in descendant infection chains include purposely un-vaccinated
individuals, a significant fraction can consist of vunerable people who had no choice but to be un-vaccinated~\cite{caplan2012free}.

%While the disease we examined in this hypothetical case study was measles, our model
%is not specific to measles, and indeed can be used for any disease that confers
%lifelong immunity upon recovery, and/or the vaccine has very high efficacy.

The model used in this initial work made several simplifying assumptions,
including homogeneous mixing of the individuals.  It also did not explicitly include
a vaccinated class, but as mentioned above, we could nevertheless use the model
to assess ANDI in the scenario of
where a portion of the individuals in a population are vaccinated with a fully effective vaccine.
The deterministic and stochastic formalisms we have developed form the basis
for a wide array of future related work,
including expanding the model to include partially effective
vaccination, age groups, 
other heterogeneities in transmission, latent periods, etc.
%The model we examined, like most compartmental disease models of its kind, also assumes that the probability distribution for the sojourn time in the infectious state is exponentially distributed~\cite{yan2010variability}; the potential effect of more realistic non-exponential probability distributions on ANDI also needs to be assessed.
While we examined measles-type outbreaks in this analysis, the modelling formalism can 
be expanded to simulate outbreaks for a wide variety of 
other vaccine-preventable diseases, including pertussis, influenza, and varicella.

We expect this work and its future derivatives will be impactful 
in informing the vaccination debate, particularly
for hesitant parents who might be swayed by the realisation that what might appear to be
a personal decision that only affects their child
actually has potential grave impacts on others in society.

%%%%%%%%%%%%%%%%%%%%%%%%%%%%%%%%%%%%%%%%%%%%%%%%%%%%%%%%%%%%%%%%%%%%%%%%%%%%
%%%%%%%%%%%%%%%%%%%%%%%%%%%%%%%%%%%%%%%%%%%%%%%%%%%%%%%%%%%%%%%%%%%%%%%%%%%%
%%%%%%%%%%%%%%%%%%%%%%%%%%%%%%%%%%%%%%%%%%%%%%%%%%%%%%%%%%%%%%%%%%%%%%%%%%%%
\bibliographystyle{unsrt}

\end{document}